\documentclass[11pt]{article}
\usepackage{amssymb,amsmath,amsfonts}
\usepackage{graphicx}
\usepackage{graphics}
\usepackage{eepic,epsfig}

\makeatletter
\@addtoreset{equation}{section}

\makeatother

\begin{document}

\title{The Casimir effect for parallel plates in Friedmann-Robertson-Walker
universe}
\author{E. R. Bezerra de Mello$^{1}$\thanks{%
E-mail: emello@fisica.ufpb.br}, A. A. Saharian$^{2}$\thanks{%
E-mail: saharian@ysu.am}, M. R. Setare$^{3,4}$\thanks{%
E-mail: rezakord@ipm.ir}\vspace{0.3cm} \\
\textit{$^{1}$Departamento de F\'{\i}sica, Universidade Federal da Para\'{\i}%
ba}\\
\textit{58.059-970, Caixa Postal 5.008, Jo\~{a}o Pessoa, PB, Brazil}\\
\textit{$^2$Department of Physics, Yerevan State University, 1 Alex
Manoogian Street,}\\
\textit{0025 Yerevan, Armenia} \\
\textit{$^{3}$Department of Science, Campus of Bijar, University of
Kurdistan, Bijar, Iran}\\
\textit{$^{4}$Research Institute for Astronomy and Astrophysics of Maragha
(RIAAM),}\\
\textit{P.O. Box 55134-441, Maragha, Iran}}
\maketitle

\begin{abstract}
We evaluate the Hadamard function, the vacuum expectation values (VEVs) of
the field squared and the energy-momentum tensor for a massive scalar field
with general curvature coupling parameter in the geometry of two parallel
plates on a spatially flat Friedmann-Robertson-Walker background with a
general scale factor. On the plates, the field operator obeys the Robin
boundary conditions with the coefficients depending on the scale factor. In
all the spatial regions, the VEVs are decomposed into the boundary-free and
boundary-induced contributions. Unlike to the problem with the Minkowski
bulk, in the region between the plates the normal stress is not homogeneous
and does not vanish in the geometry of a single plate. Near the plates, it
has different signs for accelerated and deccelerated expansions of the
universe. The VEV of the energy-momentum tensor, in addition to the diagonal
components, has a nonzero off-diagonal component describing an energy flux
along the direction normal to the boundaries. Expressions are derived for
the Casimir forces acting on the plates. Depending on the Robin coefficients
and on the vacuum state, these forces can be either attractive or repulsive.
An important difference from the corresponding result in the Minkowski bulk
is that the forces on the separate plates, in general, are different if the
corresponding Robin coefficients differ. We give the applications of general
results for the class of $\alpha $-vacua in the de Sitter bulk. It is shown
that, compared with the Bunch-Davies vacuum state, the Casimir forces for a
given $\alpha $-vacuum may change the sign.
\end{abstract}

\bigskip


\bigskip

\section{Introduction}

The investigation of quantum effects in cosmological backgrounds is among
the most important topics of quantum field theory in curved spacetime (see
\cite{Birr82B}). There are several reasons for that. Due to the high
symmetry of the background geometry, a relatively large number of problems
are exactly solvable and the corresponding results may shed light on the
influence of the gravitational field on quantum fields for more complicated
geometries. The expectation value of the energy-momentum tensor for quantum
fields may break the energy conditions appearing in the formulations of the
Hawking-Penrose singularity theorems. This expectation value appears as a
source for the gravitational field in the right-hand side of the Einstein
equations and, consequently, the quantum effects of nongravitational fields
may provide a way to solve the cosmological singularity problem. In the
inflationary phase, the quantum fluctuations of fields are responsible for
the generation of density perturbations serving as seeds for the large scale
structure formation in the universe. Currently, this mechanism is the most
popular one for the generation of cosmological structures. From the
cosmological point of view, another interesting quantum field theoretical
effect is the isotropization of the cosmological expansion as a result of
particle creation.

In a number of cosmological problems, additional boundary conditions are
imposed on the operators of quantum fields. These conditions may have
different physical origins. For example, they can be induced by nontrivial
spatial topology, by the presence of coexisting phases, or by branes in the
scenarios of the braneworld type. The boundary conditions modify the
spectrum of quantum fluctuations of fields and, as a consequence of that,
the expectation values of physical observables are changed. This is the
well-known Casimir effect first predicted by Casimir in 1948 (for reviews
see \cite{Eliz94}). In the present paper we consider the influence of the
cosmological expansion on the local characteristics of the scalar vacuum in
the geometry of two parallel plates. This type of problems for various
special cases have been considered previously. In particular, the vacuum
expectation values (VEVs) for parallel plates in background of de Sitter
spacetime were investigated in \cite{Seta01,Saha09} and \cite{Saha14} for
scalar and electromagnetic fields, respectively. The problems with spherical
and cylindrical boundaries have been discussed in \cite{Milt12,Saha15} (for
the Casimir densities in the anti-de Sitter bulk see \cite{Eliz13} and
references therein). All these investigations have been done for the de
Sitter invariant Bunch-Davies vacuum state. By using the conformal relation
between the Friedmann-Roberston-Walker (FRW) and the Rindler spacetimes, the
VEVs of the energy-momentum tensor and the Casimir forces for a conformally
coupled massless scalar field and for the electromagnetic field, in the
geometry of curved boundaries on background of FRW spacetime with negative
spatial curvature, were evaluated in \cite{Saha10} (for a special case of
the static background see also \cite{Seta09}). The electromagnetic Casimir
effect in FRW cosmologies with an arbitrary number of spatial dimensions and
with power-law scale factors has been considered in \cite{Bell13} (for the
topological Casimir densities in the corresponding models with compact
dimensions see \cite{Saha10b}).

The present paper generalizes the previous investigations in two directions.
First, we consider a spatially flat FRW spacetime with general scale factor
and, second, the Casimir effect will be investigated without specifying the
vacuum state for a scalar field. The boundary geometry consists of two
parallel plates on which the scalar field operator obeys the Robin boundary
conditions with, in general, different coefficients on the separate plates.
We consider the case when these coefficients are proportional to the scale
factor. With this assumption, closed analytical expressions are obtained for
the Hadamard function and for the VEVs of the field squared and the
energy-momentum tensor without specifying the time dependence of the scale
factor.

The paper is organized as follows. In the next section, we describe the bulk
and boundary geometries under consideration and the field content. In
section \ref{sec:Hadam}, the Hadamard function is evaluated for a massive
scalar field with general curvature coupling parameter and obeying the Robin
boundary conditions on two parallel plates. The boundary-induced
contributions are explicitly separated for both the single plate and
two-plates geometries. By using the Hadamard function, in section \ref%
{sec:VEVs} we evaluate the VEVs of the field squared and of the
energy-momentum tensor. Expressions are derived for the Casimir forces
acting on the plates. Two special cases of the general results are discussed
in section \ref{sec:Spec}. They include a conformally coupled massless
scalar field for a general scale factor and the de Sitter bulk with a
massive scalar field for the general case of the curvature coupling.
Finally, we leave for section \ref{sec:Conc} the most relevant discussion of
the results obtained.

\section{Problem formulation and the scalar modes}

\label{sec:Problem}

As a background geometry we take a spatially flat $(1+D)$-dimensional FRW
spacetime described by the line element
\begin{equation}
ds^{2}=dt^{2}-a^{2}(t)\sum_{i=1}^{D}(dx^{i})^{2}\ ,  \label{ds2}
\end{equation}%
with the scale factor $a(t)$. Defining the conformal time $\eta $ in terms
of the cosmic time $t$ by $\eta =\int dt/a(t)$, the metric tensor is
presented in a conformally flat form $g_{\mu \nu }=a^{2}(\eta )\eta _{\mu
\nu }$ with the flat spacetime metric $\eta _{\mu \nu }$. In addition to the
Hubble function $H=\dot{a}/a$, we will use the corresponding function for
the conformal time
\begin{equation}
\tilde{H}=a^{\prime }(\eta )/a(\eta )=a(\eta )H.  \label{3}
\end{equation}%
Here and in what follows, the dot specifies the derivative with respect to
the cosmic time and the prime denotes the derivative with respect to the
conformal time.

Consider a massive scalar field $\phi (x)$ non-minimally coupled to the
background. The corresponding action functional has the form
\begin{equation}
S=\frac{1}{2}\int d^{D+1}x\sqrt{|g|}\left( g^{\mu \nu }\nabla _{\mu }\phi
\nabla _{\nu }\phi -m^{2}\phi ^{2}-\xi R\phi ^{2}\right) \ ,  \label{4}
\end{equation}%
where $\nabla _{\mu }$ stands for the covariant derivative and $\xi $ is the
coupling parameter to the curvature scalar $R$. For the background geometry
under consideration one has%
\begin{equation}
R=\frac{D}{a^{2}}\left[ 2\tilde{H}^{\prime }+(D-1)\tilde{H}^{2}\right] .
\label{R}
\end{equation}%
By varying the action with respect to the field, one obtains the equation of
motion
\begin{equation}
(\nabla _{\mu }\nabla ^{\mu }+m^{2}+\xi R)\phi =0\ .  \label{5}
\end{equation}%
Additionally, we assume the presence of two flat boundaries located at $%
z\equiv x^{D}=z_{1}$ and $z=z_{2}$, $z_{2}>z_{1}$, on which the field
satisfies the Robin boundary conditions
\begin{equation}
(1+\beta _{j}^{\prime }n_{j}^{\mu }\nabla _{\mu })\phi =0,\;z=z_{j},\;j=1,2,
\label{6}
\end{equation}%
where $n_{j}^{\mu }$ is the normal to the boundary $z=z_{j}$, $n_{j\mu
}n_{j}^{\mu }=-1$. For the region between the plates, $z_{1}\leqslant
z\leqslant z_{2}$, one has $n_{j}^{\mu }=(-1)^{j-1}\delta _{D}^{\mu }/a(\eta
)$. The Robin condition is an extension of the Dirichlet and Neumann
boundary conditions and is useful for modeling the finite penetration of the
field into the boundary with the skin-depth parameter related to the
coefficient $\beta _{j}^{\prime }$ \cite{Most85}. This type of boundary
conditions naturally arise for bulk fields in braneworld models. In the
discussion below we will consider a class of boundary conditions for which $%
\beta _{j}^{\prime }=\beta _{j}a(\eta )$ with $\beta _{j}$, $j=1,2$, being
constants. This corresponds to the physical situation when for an expanding
bulk the penetration length to the boundary is increasing as well. In this
special case, for the region between the plates, the boundary conditions are
rewritten as
\begin{equation}
(1+(-1)^{j-1}\beta _{j}\partial _{z})\phi =0,\;z=z_{j}.  \label{7}
\end{equation}%
As it will be shown below, the corresponding Casimir problem is exactly
solvable for general case of the scale factor $a(\eta )$.

We are interested in the changes of the VEVs of the field squared and the
energy-momentum tensor induced by the imposition of the boundary conditions (%
\ref{7}). The VEVs of physical observables, quadratic in the field operator,
are expressed in terms of the sums over a complete set of solutions to the
field equation (\ref{5}) obeying the boundary conditions. In accordance with
the geometry of the problem, for the corresponding mode functions we will
use the ansatz
\begin{equation}
\phi (x)=f(\eta )e^{i\mathbf{k}\cdot \mathbf{x}_{\parallel }}h(z)\ ,\
\mathbf{k}=(k^{1},k^{2},\ldots ,k^{D-1})\ ,\ \mathbf{x}_{\parallel
}=(x^{1},x^{2},\ldots ,x^{D-1}).  \label{ansatz}
\end{equation}%
Substituting into the field equation (\ref{5}), we obtain two differential
equations:
\begin{equation}
h^{\prime \prime }(z)=-\lambda ^{2}h(z)\   \label{h}
\end{equation}%
and
\begin{equation}
f^{\prime \prime }(\eta )+(D-1)\tilde{H}f^{\prime }(\eta )+\left[ \gamma
^{2}+a^{2}\left( m^{2}+\xi R\right) \right] f(\eta )=0,  \label{f1}
\end{equation}%
with $\gamma =\sqrt{\lambda ^{2}+{k}^{2}}$ and $k=|\mathbf{k}|$. In
particular, from the equation (\ref{f1}) it follows that%
\begin{equation}
\left\{ a^{D-1}\left[ f^{\ast }(\eta )f^{\prime }(\eta )-f(\eta )f^{\ast
\prime }(\eta )\right] \right\} ^{\prime }=0,  \label{cond}
\end{equation}%
where the star stands for the complex conjugate. Note that, introducing the
function $g(\eta )=a^{(D-1)/2}f(\eta )$, the equation (\ref{f1}) is written
in the form%
\begin{equation}
g^{\prime \prime }(\eta )+\left\{ \gamma ^{2}+m^{2}a^{2}+D\left( \xi -\xi
_{D}\right) \left[ 2\tilde{H}^{\prime }+(D-1)\tilde{H}^{2}\right] \right\}
g(\eta )=0,  \label{g1}
\end{equation}%
where $\xi _{D}=(D-1)/(4D)$ is the curvature coupling parameter for a
conformally coupled scalar field.

The solution for (\ref{h}) that obeys the boundary condition on the plate $%
z=z_{j}$ reads,
\begin{equation}
h(z)=\cos \left[ \lambda \left( z-z_{j}\right) +\alpha _{j}(\lambda )\right]
\ ,  \label{h1}
\end{equation}%
with the function $\alpha _{j}(\lambda )$ defined by the relation
\begin{equation}
e^{2i\alpha _{j}(\lambda )}=\frac{i\lambda \beta _{j}(-1)^{j}+1}{i\lambda
\beta _{j}(-1)^{j}-1}.  \label{alfj}
\end{equation}%
From the boundary condition on the second plate it follows that the quantum
number $\lambda $ obeys the restriction condition
\begin{equation}
(1-b_{1}b_{2}u^{2})\sin u-(b_{1}+b_{2})u\cos u\ =0,  \label{rest}
\end{equation}%
where%
\begin{equation}
u=\lambda z_{0},\;b_{j}=\beta _{j}/z_{0},\;z_{0}=z_{2}-z_{1}.  \label{bj}
\end{equation}%
Note that the eigenvalue equation (\ref{rest}) coincides with the
corresponding equation for parallel plates in the Minkowski bulk \cite{RS02}%
. We will denote the solutions of the transcendental equation (\ref{rest})
by $u=u_{n}$, $n=1,2,\ldots $. For the eigenvalues of the quantum number $%
\lambda $ one has $\lambda =\lambda _{n}=u_{n}/z_{0}$.

So, for the complete set of solutions one has $\{\phi _{\sigma
}^{(+)}(x),\phi _{\sigma }^{(-)}(x)\}$, where
\begin{equation}
\phi _{\sigma }^{(+)}(x)=C_{\sigma }f(\eta ,\gamma )e^{i\mathbf{k}\cdot
\mathbf{x}_{\parallel }}\cos \left[ \lambda _{n}\left( z-z_{j}\right)
+\alpha _{j}(\lambda _{n})\right] \ ,  \label{phi2}
\end{equation}%
$\phi _{\sigma }^{(-)}(x)=\phi _{\sigma }^{(+)\ast }(x)$, with $C_{\sigma }$
being a normalization constant and $\sigma =(n,\mathbf{k})$ representing the
set of quantum numbers specifying the modes. In (\ref{phi2}), the dependence
of the function $f$ on $\gamma $ is explicitly displayed.

In accordance with (\ref{cond}), we will normalize the function $f(\eta
,\gamma )$ by the condition%
\begin{equation}
f(\eta ,\gamma )\partial _{\eta }f^{\ast }(\eta ,\gamma )-f^{\ast }(\eta
,\gamma )\partial _{\eta }f(\eta ,\gamma )=ia^{1-D}.  \label{cond2}
\end{equation}%
With this normalization, the constant $C_{\sigma }$ is determined from the
standard orthonormalization condition for the Klein-Gordon equation:
\begin{equation}
\int \ d^{D}x\sqrt{|g|}g^{00}\left[ \phi _{\sigma }(x)\partial _{\eta }\phi
_{\sigma ^{\prime }}^{\ast }(x)-\phi _{\sigma ^{\prime }}^{\ast }(x)\partial
_{\eta }\phi _{\sigma }(x)\right] =i\delta \left( \mathbf{k}-\mathbf{k}%
^{\prime }\right) \delta _{nn^{\prime }}.  \label{nor}
\end{equation}%
By taking into account (\ref{cond2}), one gets%
\begin{equation}
\left\vert C_{\sigma }\right\vert ^{2}=\frac{2}{\left( 2\pi \right)
^{D-1}z_{0}}\left\{ 1+\frac{\sin u_{n}}{u_{n}}\cos [u_{n}+2\tilde{\alpha}%
_{j}(u_{n})]\right\} ^{-1}\ ,  \label{Csig}
\end{equation}%
where the function $\tilde{\alpha}_{j}(u)$ is defined in accordance with
\begin{equation}
e^{2i\tilde{\alpha}_{j}(u)}=\frac{iub_{j}-1}{iub_{j}+1},  \label{alftild}
\end{equation}%
for $j=1,2$.

Note that the mode functions (\ref{phi2}) are not yet completely fixed. The
function $f(\eta ,\gamma )$ is a linear combination of two linearly
independent solutions of the equation (\ref{f1}). One of the coefficients is
fixed (up to a phase) by the condition (\ref{cond2}). Among the most
important steps in the construction of a quantum field theory in a fixed
classical gravitational background is the choice of the vacuum state $%
|0\rangle $. Different choices of the second coefficient in the linear
combination for the function $f(\eta ,\gamma )$ correspond to different
choices of the vacuum state. An additional condition could be the
requirement of the smooth transition to the standard Minkowskian vacuum in
the limit of slow expansion. This point will be discussed below for an
example of de Sitter bulk.

In the limit of small wavelengths, $\gamma \gg ma,\sqrt{|\tilde{H}^{\prime }|%
},\tilde{H}$, the general solution of the equation (\ref{g1}) is a linear
combination of the functions $e^{i\gamma \eta }$ and $e^{-i\gamma \eta }$.
For the modes which satisfy the adiabatic condition (for the adiabatic
condition see \cite{Birr82B}) one takes $g(\eta )\sim e^{-i\gamma \eta }$
and the function $f(\eta ,\gamma )$, normalized by the condition (\ref{cond2}%
) has the small wavelength asymptotic behavior%
\begin{equation}
f(\eta ,\gamma )\approx a^{(1-D)/2}\frac{e^{-i\gamma \eta }}{\sqrt{2\gamma }}%
.  \label{fadiab}
\end{equation}%
In the limit of slow expansion, these modes approach the positive energy
solutions for a scalar field in Minkowski spacetime. The condition on the
wavelength, written in terms of the cosmic time $t$, is in the form $\gamma
/a\gg m,\sqrt{|\dot{H}|},H$. Note that the condition (\ref{fadiab}) does not
specify the vacuum state uniquely (for the discussion of related
uncertainties in the inflationary predictions of the curvature perturbations
see, for instance, \cite{Chun03}).

\section{Hadamard function}

\label{sec:Hadam}

Given the complete set of modes, we can evaluate the two-point functions. We
consider a free field theory (the only interaction is with the background
gravitational field) and all the information about the vacuum state is
encoded in two-point functions. As such we take the Hadamard function $%
G(x,x^{\prime })=$ $\langle 0|\phi (x)\phi (x^{\prime })+\phi (x^{\prime
})\phi (x)|0\rangle $ with the mode sum formula%
\begin{equation}
G(x,x^{\prime })=\int d\mathbf{k}\,\sum_{n=1}^{\infty }\sum_{s=\pm }\phi
_{\sigma }^{(s)}(x)\phi _{\sigma }^{(s)\ast }(x^{\prime }).  \label{Had}
\end{equation}%
Substituting the mode functions (\ref{phi2}) one gets the representation%
\begin{eqnarray}
G(x,x^{\prime }) &=&\frac{2}{z_{0}}\int d\mathbf{k}\,\frac{e^{i\mathbf{k}%
\cdot \Delta \mathbf{x}_{\parallel }}}{\left( 2\pi \right) ^{D-1}}%
\sum_{n=1}^{\infty }u_{n}w(\eta ,\eta ^{\prime },\gamma _{n})  \notag \\
&&\times \frac{\cos \left[ \lambda _{n}(z-z_{j})+\alpha _{j}(\lambda _{n})%
\right] \cos \left[ \lambda _{n}(z^{\prime }-z_{j})+\alpha _{j}(\lambda _{n})%
\right] }{u_{n}+\sin u_{n}\cos [u_{n}+2\tilde{\alpha}_{j}(\lambda _{n})]},
\label{Had1}
\end{eqnarray}%
with $\Delta \mathbf{x}_{\parallel }=\mathbf{x}_{\parallel }-\mathbf{x}%
_{\parallel }^{\prime }$, $\gamma _{n}=\sqrt{\lambda _{n}^{2}+{k}^{2}}$, and%
\begin{equation}
w(\eta ,\eta ^{\prime },\gamma )=f(\eta ,\gamma )f^{\ast }(\eta ^{\prime
},\gamma )+f^{\ast }(\eta ,\gamma )f(\eta ^{\prime },\gamma ).  \label{w}
\end{equation}

In (\ref{Had1}), $\lambda _{n}=u_{n}/z_{0}$ and the eigenvalues $u_{n}$ are
given implicitly, as solutions of (\ref{rest}). Related to this, the
representation (\ref{Had1}) is not convenient for the evaluation of the
VEVs. For the further transformation, we apply to the series over $n$ the
summation formula \cite{RS02,SahaBook}:%
\begin{eqnarray}
\sum_{n=1}^{\infty }\frac{\pi u_{n}s(u_{n})}{u_{n}+\sin u_{n}\cos [u_{n}+2%
\tilde{\alpha}_{j}(\lambda _{n})]} &=&-\frac{\pi s(0)/2}{1-b_{2}-b_{1}}%
+\int_{0}^{\infty }du\,s(u)  \notag \\
&&+i\int_{0}^{\infty }du\frac{s(iu)-s(-iu)}{c_{1}(u)c_{2}(u)e^{2u}-1},
\label{sumfor}
\end{eqnarray}%
where the notation%
\begin{equation}
c_{j}(u)=\frac{b_{j}u-1}{b_{j}u+1}  \label{cj}
\end{equation}%
is introduced. In (\ref{sumfor}), it is assumed that the function $s(u)$
obeys the condition $|s(u)|<\epsilon (x)e^{c|y|}$ for $|u|\rightarrow \infty
$, where $u=x+iy$, $c<2$, and $\epsilon (x)\rightarrow 0$ for $x\rightarrow
\infty $. As the function $s(u)$ we take%
\begin{equation}
s(u)=\left\{ \cos \left( uz_{-}/z_{0}\right) +\cos [u\left(
z_{+}-2z_{j}\right) /z_{0}+2\alpha _{j}(u/z_{0})]\right\} w(\eta ,\eta
^{\prime },\sqrt{u^{2}/z_{0}^{2}+k^{2}}),  \label{su}
\end{equation}%
with%
\begin{equation}
z_{\pm }=z\pm z^{\prime }.  \label{zpm}
\end{equation}

After the application of (\ref{sumfor}), the Hadamard function (\ref{Had1})
is decomposed as%
\begin{eqnarray}
G(x,x^{\prime }) &=&G_{j}(x,x^{\prime })+\frac{1}{z_{0}}\int d\mathbf{k}\,%
\frac{e^{i\mathbf{k}\cdot \Delta \mathbf{x}_{\parallel }}}{\left( 2\pi
\right) ^{D}}\int_{0}^{\infty }du\frac{W(\eta ,\eta ^{\prime },u,k)}{%
c_{1}(u)c_{2}(u)e^{2u}-1}  \notag \\
&&\times \left[ 2\cosh \left( uz_{-}/z_{0}\right) +c_{j}\left( u\right)
e^{u|z_{+}-2z_{j}|/z_{0}}+\frac{e^{-u|z_{+}-2z_{j}|/z_{0}}}{c_{j}\left(
u\right) }\right] ,  \label{Had2}
\end{eqnarray}%
where%
\begin{equation}
W(\eta ,\eta ^{\prime },u,k)=i\left[ w(\eta ,\eta ^{\prime },\sqrt{%
(iu)^{2}/z_{0}^{2}+k^{2}})-w(\eta ,\eta ^{\prime },\sqrt{%
(-iu)^{2}/z_{0}^{2}+k^{2}})\right] .  \label{Wf}
\end{equation}%
The part%
\begin{equation}
G_{j}(x,x^{\prime })=G_{0}(x,x^{\prime })+2\int d\mathbf{k}\,\frac{e^{i%
\mathbf{k}\cdot \Delta \mathbf{x}_{\parallel }}}{\left( 2\pi \right) ^{D}}%
\int_{0}^{\infty }dy\,\cos [y\left( z_{+}-2z_{j}\right) +2\alpha
_{j}(y)]w(\eta ,\eta ^{\prime },\sqrt{y^{2}+k^{2}}),  \label{Gj}
\end{equation}%
comes from the first integral in the right-hand side of (\ref{sumfor}) and
it presents the Hadamard function for the geometry of a single plate at $%
z=z_{j}$ when the second plate is absent. In (\ref{Gj}), the contribution%
\begin{equation}
G_{0}(x,x^{\prime })=\int d\mathbf{k}_{D}\,\frac{e^{i\mathbf{k}_{D}\cdot
\Delta \mathbf{x}}}{\left( 2\pi \right) ^{D}}w(\eta ,\eta ^{\prime },|%
\mathbf{k}_{D}|),  \label{G0}
\end{equation}%
with $\mathbf{x}=(x^{1},x^{2},\ldots ,x^{D})$, $\mathbf{k}%
_{D}=(k^{1},k^{2},\ldots ,k^{D})$, is the Hadamard function in the
boundary-free geometry. The second term in the right-hand side of (\ref{Gj})
is induced by the boundary at $z=z_{j}$. Consequently, the last term in (\ref%
{Had2}) is interpreted as the contribution when one adds the second boundary
in the problem with a single boundary at $z=z_{j}$.

For the further transformation of the boundary-induced contribution in (\ref%
{Gj}) we present the cosine function in terms of the exponentials and rotate
the integration contour over $y$ by the angles $\pi /2$ and $-\pi /2$ for
the parts with the functions $e^{iy|z_{+}-2z_{j}|}$ and $%
e^{-iy|z_{+}-2z_{j}|}$, respectively. As a result, for the Hadamard function
in the geometry of a single plate at $z=z_{j}$ we get%
\begin{equation}
G_{j}(x,x^{\prime })=G_{0}(x,x^{\prime })+\int d\mathbf{k}\,\frac{e^{i%
\mathbf{k}\cdot \Delta \mathbf{x}_{\parallel }}}{\left( 2\pi \right) ^{D}}%
\int_{0}^{\infty }dy\,\frac{\beta _{j}y+1}{\beta _{j}y-1}e^{-y\left\vert
z_{+}-2z_{j}\right\vert }W(\eta ,\eta ^{\prime },yz_{0},k).  \label{Gj1}
\end{equation}%
Substituting this representation into (\ref{Had2}), the Hadamard function in
the region between two plates is presented in the form%
\begin{eqnarray}
G(x,x^{\prime }) &=&G_{0}(x,x^{\prime })+\frac{1}{z_{0}}\int d\mathbf{k}\,%
\frac{e^{i\mathbf{k}\cdot \Delta \mathbf{x}_{\parallel }}}{\left( 2\pi
\right) ^{D}}\int_{0}^{\infty }du\frac{W(\eta ,\eta ^{\prime },u,k)}{%
c_{1}(u)c_{2}(u)e^{2u}-1}  \notag \\
&&\times \left[ 2\cosh \left( uz_{-}/z_{0}\right)
+\sum_{j=1,2}c_{j}(u)e^{u|z_{+}-2z_{j}|/z_{0}}\right] .  \label{Had3}
\end{eqnarray}%
This expression can be further simplified by integrating over the angular
part of $\mathbf{k}$. The corresponding integral is expressed in terms of
the Bessel function. In the regions $z<z_{1}$ and $z>z_{2}$, the Hadamard
function is given by (\ref{Gj1}) with $j=1$ and $j=2$, respectively. Note
that the dependence on the mass of the field appears in (\ref{Had3}) through
the function $f(\eta ,\gamma )$. The equation (\ref{f1}) for the latter
contains the mass as a parameter.

\section{VEVs and the Casimir force}

\label{sec:VEVs}

Having the two point function we can evaluate the VEVs of local physical
observables bilinear in the field operator.

\subsection{Field squared}

We start with the VEV of the field squared. The latter is obtained from the
Hadamard function in the coincidence limit of the arguments. Of course, this
limit is divergent and a renormalization procedure is required. An important
point is that we have separated the part of the Hadamard function
corresponding to the boundary-free geometry. For points away from boundaries
the divergences are contained in this part only and the remaining
boundary-induced contribution is finite in the coincidence limit. As a
consequence, the renormalization is reduced to that for the VEVs in the
boundary-free geometry. These VEVs are well investigated in the literature
and in the following we will focus on the boundary-induced effects.

Taking the limit $x^{\prime }\rightarrow x$ in (\ref{Had3}), for the VEV\ of
the field squared, $\langle 0|\phi ^{2}|0\rangle \equiv \langle \phi
^{2}\rangle $, in the region between the plates we get:%
\begin{equation}
\langle \phi ^{2}\rangle =\langle \phi ^{2}\rangle _{0}+\frac{A_{D}}{z_{0}}%
\int_{0}^{\infty }dk\,k^{D-2}\,\int_{0}^{\infty }du\frac{W(\eta ,\eta ,u,k)}{%
c_{1}(u)c_{2}(u)e^{2u}-1}\left[ 2+\sum_{j=1,2}c_{j}(u)e^{2u|z-z_{j}|/z_{0}}%
\right] ,  \label{phi2n}
\end{equation}%
where%
\begin{equation}
A_{D}=\frac{2^{-D}\pi ^{-(D+1)/2}}{\Gamma ((D-1)/2)},  \label{AD}
\end{equation}%
and $\langle \phi ^{2}\rangle _{0}$ is the renormalized VEV in the
boundary-free geometry. The latter does not depend on the spatial point. In
the regions $z<z_{1}$ and $z>z_{2}$, the VEVs\ are obtained from (\ref{Gj1}):%
\begin{equation}
\langle \phi ^{2}\rangle _{j}=\langle \phi ^{2}\rangle
_{0}+A_{D}\int_{0}^{\infty }dk\,k^{D-2}\int_{0}^{\infty }dy\,\frac{\beta
_{j}y+1}{\beta _{j}y-1}e^{-2y\left\vert z-z_{j}\right\vert }W(\eta ,\eta
,yz_{0},k),  \label{phi22}
\end{equation}%
with $j=1$ and $j=2$, respectively.

Alternative expressions are obtained by taking into account that $W(\eta
,\eta ,u,k)=0$ for $u<z_{0}k$ and
\begin{equation}
W(\eta ,\eta ,u,k)=U(\eta ,\sqrt{u^{2}-z_{0}^{2}k^{2}}),  \label{W2}
\end{equation}%
for $u>z_{0}k$, where%
\begin{equation}
U(\eta ,z_{0}x)=i\left[ w(\eta ,\eta ,ix)-w(\eta ,\eta ,-ix)\right] .
\label{Un}
\end{equation}%
By using the relation%
\begin{equation}
\int_{0}^{\infty }dx\,x^{n-1}\int_{x}^{\infty }du\,\,f_{1}(u)f_{2}(\sqrt{%
u^{2}-x^{2}})=\int_{0}^{\infty
}du\,u^{n}\,f_{1}(u)\int_{0}^{1}ds\,s(1-s^{2})^{n/2-1}f_{2}(us),
\label{Rel2}
\end{equation}%
the VEV\ of the field squared in the region between the plates is presented
as%
\begin{equation}
\langle \phi ^{2}\rangle =\langle \phi ^{2}\rangle _{0}+\frac{A_{D}}{%
z_{0}^{D}}\int_{0}^{\infty }du\,u^{D-1}\,Z(\eta ,u)\frac{2+%
\sum_{j=1,2}c_{j}(u)e^{2u|z-z_{j}|/z_{0}}}{c_{1}(u)c_{2}(u)e^{2u}-1},
\label{phi23}
\end{equation}%
with the notation%
\begin{equation}
Z(\eta ,u)=\int_{0}^{1}ds\,s(1-s^{2})^{(D-3)/2}U(\eta ,us).  \label{Z}
\end{equation}%
In a similar way, for the regions $z<z_{1}$ and $z>z_{2}$ from (\ref{phi22})
we get%
\begin{equation}
\langle \phi ^{2}\rangle _{j}=\langle \phi ^{2}\rangle
_{0}+A_{D}\int_{0}^{\infty }dy\,y^{D-1}Z(\eta ,yz_{0})\frac{\beta _{j}y+1}{%
\beta _{j}y-1}e^{-2y\left\vert z-z_{j}\right\vert }.  \label{phi24}
\end{equation}%
The information on the background geometry is encoded in the function $%
Z(\eta ,u)$.

For the modes which satisfy the adiabatic condition in the limit of small
wavelengths, one has the asymptotic condition (\ref{fadiab}). In this case
we can obtain simple asymptotic expressions for the VEV of the field squared
near the boundaries. From (\ref{fadiab}) it follows that for large $x$ one
has $U(\eta ,x)\approx 2z_{0}a^{1-D}/x$ and, hence, for the function $Z(\eta
,u)$ we get%
\begin{equation}
Z(\eta ,u)\approx z_{0}\frac{\sqrt{\pi }\Gamma ((D-1)/2)}{\Gamma
(D/2)a^{D-1}u},  \label{Zas}
\end{equation}%
for $u\gg 1$. In order to find the asymptotic behavior of the VEV (\ref%
{phi23}) near the boundary $z=z_{j}$, we note that in this region the
dominant contribution to the integral comes from large values of $u$. By
using (\ref{Zas}), to the leading order one gets%
\begin{equation}
\langle \phi ^{2}\rangle \approx \frac{\left( 1-2\delta _{0\beta
_{j}}\right) \Gamma ((D-1)/2)}{(4\pi )^{(D+1)/2}\left( a|z-z_{j}|\right)
^{D-1}}.  \label{phi2near}
\end{equation}%
This leading term comes from the single plate part (\ref{phi24}) and
coincides with that for the plate in Minkowski bulk with the distance from
the plate $|z-z_{j}|$ replaced by the proper distance $a(\eta )|z-z_{j}|$
for a fixed $\eta $. The latter property is natural, because, due to the
adiabatic condition, the influence of the background gravitational field on
the modes with small wavelengths is weak and in the region near the plates
the main contribution to the VEVs comes from those modes.

The regularization procedure we have employed for the evaluation of the VEV
of the field squared is based on the point-splitting technique with
combination with the summation formula (\ref{sumfor}). Instead, we could
start directly from the divergent expression $\langle \phi ^{2}\rangle
=G(x,x)/2$ with $G(x,x)$, obtained from (\ref{Had1}) in the coincidence
limit. In that expression the integration over the angular part of $\mathbf{k%
}$ is trivial. For the regularization we can introduce a cutoff function $%
F(\alpha ,\gamma _{n})$ with a regularization parameter $\alpha $, $%
F(0,\gamma _{n})=1$ (for example, $F(x)=e^{-\alpha x}$, $\alpha >0$), and
then apply Eq. (\ref{sumfor}) for the summation over $n$. For points outside
the plates, the boundary-induced contribution in the VEV of the field
squared is finite and the limit $\alpha \rightarrow 0$ can be put directly.
The corresponding result for the boundary-induced part will coincide with
the last term in Eq. (\ref{phi2n}). Another regularization procedure for the
VEVs is the local zeta function technique (see, for instance, \cite{More97}
and references therein). In the formula for the VEV $\langle \phi
^{2}\rangle $ we can introduce the factor $\gamma _{n}^{-s}$. For sufficiently
large $\mathrm{Re\,}s$ the corresponding expression is finite. For the
analytic continuation to the physical value $s=0$ we can again use Eq. (\ref%
{sumfor}). Now, in the generalized Abel-Plana formula the singular points $%
\pm ik$ should be excluded by small semicircles in the right-half plane. For
points away from the plates, the additional contributions to the
boundary-induced parts coming from the corresponding integrals vanish in the
limit $s\rightarrow 0$. The boundary-induced parts are finite for $s=0$ and
this value can be directly substituted in the integrand with the results in
agreement with those we have displayed before.

\subsection{Energy-momentum tensor}

Another important characteristic of the vacuum state is the VEV of the
energy-momentum tensor, $\langle 0|T_{\mu \nu }|0\rangle \equiv \langle
T_{\mu \nu }\rangle $. Given the Hadamard function and the VEV\ of the field
squared, it is evaluated by using the formula%
\begin{equation}
\langle T_{\mu \nu }\rangle =\frac{1}{2}\lim_{x^{\prime }\rightarrow
x}\partial _{\mu }\partial _{\nu }^{\prime }G(x,x^{\prime })+\left[ \left(
\xi -1/4\right) g_{\mu \nu }\nabla _{l}\nabla ^{l}-\xi \nabla _{\mu }\nabla
_{\nu }-\xi R_{\mu \nu }\right] \langle \phi ^{2}\rangle ,  \label{emtvev1}
\end{equation}%
where for the Ricci tensor one has%
\begin{equation}
R_{00}=D\tilde{H}^{\prime },\;R_{ii}=-\tilde{H}^{\prime }-(D-1)\tilde{H}^{2},
\label{Rik}
\end{equation}%
with $i=1,2,\ldots ,D$, and the off-diagonal components vanish.

By taking into account the expression (\ref{Had3}) for the Hadamard
function, the diagonal components of the vacuum energy-momentum tensor are
presented as (no summation over $\nu $)%
\begin{eqnarray}
\langle T_{\nu }^{\nu }\rangle &=&\langle T_{\nu }^{\nu }\rangle _{0}+\frac{%
A_{D}}{z_{0}a^{2}}\int_{0}^{\infty }dk\,k^{D-2}\int_{0}^{\infty }du  \notag
\\
&&\times \frac{2F_{\nu }(\eta ,u,k)+G_{\nu }(\eta
,u,k)\sum_{j=1,2}c_{j}(u)e^{2u|z-z_{j}|/z_{0}}}{c_{1}(u)c_{2}(u)e^{2u}-1},
\label{Tnu}
\end{eqnarray}%
where $\langle T_{\nu }^{\nu }\rangle _{0}$ is the corresponding VEV in the
boundary-free geometry. In (\ref{Tnu}), we have defined the functions%
\begin{eqnarray}
F_{0}(\eta ,u,k) &=&W_{0}(\eta ,u,k)-\hat{P}W(\eta ,\eta ,u,k),  \notag \\
F_{l}(\eta ,u,k) &=&\left[ -\hat{P}_{1}-k^{2}/(D-1)\right] W(\eta ,\eta
,u,k),  \notag \\
F_{D}(\eta ,u,k) &=&\left[ -\hat{P}_{1}+\left( u/z_{0}\right) ^{2}\right]
W(\eta ,\eta ,u,k),  \label{FD}
\end{eqnarray}%
for $l=1,\ldots ,D-1$, and%
\begin{equation}
G_{\nu }(\eta ,u,k)=F_{\nu }(\eta ,u,k)+b_{\nu }\left( u/z_{0}\right)
^{2}W(\eta ,\eta ,u,k),  \label{Gnu}
\end{equation}%
where $b_{\nu }=1-4\xi $ for$\;\nu \neq D$, $b_{D}=-1$. For the operators in
(\ref{FD}) one has%
\begin{eqnarray}
\hat{P} &=&(1/4)\partial _{\eta }^{2}-D\left( \xi -\xi _{D}\right) \tilde{H}%
\partial _{\eta }+D\xi \tilde{H}^{\prime },  \notag \\
\hat{P}_{1} &=&\left( \frac{1}{4}-\xi \right) \partial _{\eta }^{2}+\left[
\frac{D-1}{4}-(D-2)\xi \right] \tilde{H}\partial _{\eta }+\xi \left[ \tilde{H%
}^{\prime }+(D-1)\tilde{H}^{2}\right] ,  \label{P12}
\end{eqnarray}%
and%
\begin{equation}
W_{0}(\eta ,u,k)=\lim_{\eta ^{\prime }\rightarrow \eta }\partial _{\eta
}\partial _{\eta ^{\prime }}W(\eta ,\eta ^{\prime },u,k).  \label{W0}
\end{equation}%
Due to the homogeneity of the background spacetime, the boundary-free
contribution $\langle T_{\nu }^{\nu }\rangle _{0}$ to (\ref{Tnu}) does not
depend on the spatial point (for the VEV\ of the energy-momentum tensor in
boundary-free FRW cosmologies see, for instance, \cite{Birr82B} and Refs.
\cite{Ande99} for more recent discussions).

By using the equation (\ref{f1}), it can be seen that%
\begin{equation}
W_{0}(\eta ,u,k)=\left[ \frac{1}{2}\partial _{\eta }^{2}+\frac{D-1}{2}\tilde{%
H}\partial _{\eta }+k^{2}-\frac{u^{2}}{z_{0}^{2}}+a^{2}\left( m^{2}+\xi
R\right) \right] W(\eta ,\eta ,u,k).  \label{W01}
\end{equation}%
Substituting this into (\ref{FD}), we get an alternative expression for the
function $F_{0}(\eta ,u,k)$:%
\begin{equation}
F_{0}(\eta ,u,k)=\left( \hat{P}_{0}+k^{2}-u^{2}/z_{0}^{2}\right) W(\eta
,\eta ,u,k),  \label{F0}
\end{equation}%
with the operator%
\begin{equation}
\hat{P}_{0}=\frac{1}{4}\partial _{\eta }^{2}+D\left( \xi +\xi _{D}\right)
\tilde{H}\partial _{\eta }+a^{2}m^{2}+D\xi \left[ \tilde{H}^{\prime }+(D-1)%
\tilde{H}^{2}\right] .  \label{P0}
\end{equation}%
Note that one has $G_{D}(\eta ,u,k)=-\hat{P}_{1}W(\eta ,\eta ,u,k)$ and this
function vanishes for the Minkowski bulk. Hence, in the latter geometry the
normal stress is homogeneous. In general, this is not the case for the FRW
background.

The problem under consideration is inhomogeneous along the $t$- and $z$%
-directions. As a consequence of that, in addition to the diagonal
components, the vacuum energy-momentum tensor has a nonzero off-diagonal
component%
\begin{equation}
\langle T_{0}^{D}\rangle =-\frac{A_{D}}{z_{0}a^{2}}\int_{0}^{\infty
}dk\,k^{D-2}\int_{0}^{\infty }du\,u\frac{%
\sum_{j=1,2}c_{j}(u)(-1)^{j-1}e^{2u|z-z_{j}|/z_{0}}}{c_{1}(u)c_{2}(u)e^{2u}-1%
}G_{0D}(\eta ,u,k),  \label{T0D1}
\end{equation}%
with the notation%
\begin{equation}
G_{0D}(\eta ,u,k)=\left[ \left( 1/2-2\xi \right) \partial _{\eta }+2\xi
\tilde{H}\right] W(\eta ,\eta ,u,k).  \label{G0D}
\end{equation}%
This corresponds to the energy flux along the direction perpendicular to the
plates. If the Robin coefficients for the boundaries are the same, one has $%
c_{1}(u)=c_{2}(u)$. In this special case, the energy flux $\langle
T_{0}^{D}\rangle $ vanishes at $z=(z_{1}+z_{2})/2$ and has opposite signs in
the regions $z<(z_{1}+z_{2})/2$ and $z>(z_{1}+z_{2})/2$. Note that we have
the relation%
\begin{equation}
\partial _{\eta }\left( a^{D-1}G_{0D}(\eta ,u,k)\right) =-a^{D+1}G_{D}(\eta
,u,k),  \label{relG}
\end{equation}%
between the functions in the expressions for the normal stress and the
energy flux.

In the regions $z<z_{1}$ and $z>z_{2}$, for the VEV\ of the energy-momentum
tensor one has (no summation over $\nu $)%
\begin{eqnarray}
\langle T_{\nu }^{\nu }\rangle _{j} &=&\langle T_{\nu }^{\nu }\rangle _{0}+%
\frac{A_{D}}{a^{2}}\int_{0}^{\infty }dk\,k^{D-2}\int_{0}^{\infty }dy\,\frac{%
\beta _{j}y+1}{\beta _{j}y-1}\frac{G_{\nu }(\eta ,yz_{0},k)}{e^{2y|z-z_{j}|}}%
,  \notag \\
\langle T_{0}^{D}\rangle _{j} &=&-\frac{\left( -1\right) ^{j}A_{D}}{a^{2}}%
\int_{0}^{\infty }dk\,k^{D-2}\int_{0}^{\infty }dy\,y\frac{\beta _{j}y+1}{%
\beta _{j}y-1}\frac{G_{0D}(\eta ,yz_{0},k)}{e^{2y|z-z_{j}|}}.  \label{T0D1pl}
\end{eqnarray}%
with $j=1$ and $j=2$, respectively.

By taking into account that%
\begin{equation}
\sum_{\nu =0}^{D}F_{\nu }(\eta ,u,k)=\left\{ D\left( \xi -\xi _{D}\right)
\left[ \partial _{\eta }^{2}+\left( D-1\right) \tilde{H}\partial _{\eta }%
\right] +a^{2}m^{2}\right\} W(\eta ,\eta ,u,k),  \label{sumF}
\end{equation}%
it can be explicitly checked that the boundary-induced contributions in (\ref%
{Tnu}) and (\ref{T0D1pl}), $\langle T_{\nu }^{\nu }\rangle _{\mathrm{b}%
}=\langle T_{\nu }^{\nu }\rangle -\langle T_{\nu }^{\nu }\rangle _{0}$, obey
the trace relation%
\begin{equation}
\langle T_{\mu }^{\mu }\rangle _{\mathrm{b}}=\left[ D(\xi -\xi _{D})\nabla
_{\mu }\nabla ^{\mu }+m^{2}\right] \langle \phi ^{2}\rangle _{\mathrm{b}},
\label{Trace}
\end{equation}%
where $\langle \phi ^{2}\rangle _{\mathrm{b}}=\langle \phi ^{2}\rangle
-\langle \phi ^{2}\rangle _{0}$ is the boundary-induced part in the VEV of
the field squared. For a conformally coupled massless field the
boundary-induced contribution in the VEV of the energy-momentum tensor is
traceless. The trace anomaly is contained in the boundary-free part only. As
an additional check, we can see that the boundary-induced VEVs satisfy the
covariant conservation equation $\nabla _{\mu }\langle T_{\nu }^{\mu
}\rangle _{\mathrm{b}}=0$. For the geometry under consideration it is
reduced to the following two equations%
\begin{eqnarray}
\frac{1}{a^{D+1}}\partial _{\eta }\left( a^{D+1}\langle T_{0}^{0}\rangle _{%
\mathrm{b}}\right) +\partial _{z}\langle T_{0}^{D}\rangle _{\mathrm{b}}-%
\tilde{H}\langle T_{\mu }^{\mu }\rangle _{\mathrm{b}} &=&0,  \notag \\
\partial _{z}\langle T_{D}^{D}\rangle _{\mathrm{b}}-\frac{1}{a^{D+1}}%
\partial _{\eta }\left( a^{D+1}\langle T_{0}^{D}\rangle _{\mathrm{b}}\right)
&=&0.  \label{ConsEq}
\end{eqnarray}%
In particular, the second equation directly follows from the relation (\ref%
{relG}). This equation shows that the inhomogeneity of the normal stress is
related to the nonzero energy flux along the direction normal to the plates.

Equivalent representations for the VEVs of the energy-momentum tensor are
obtained by using the relation (\ref{Rel2}). In the way similar to that we
have used for the VEV of the field squared, for the diagonal components one
gets (no summation over $\nu $)%
\begin{eqnarray}
\langle T_{\nu }^{\nu }\rangle &=&\langle T_{\nu }^{\nu }\rangle _{0}+\frac{%
A_{D}}{z_{0}^{D}a^{2}}\int_{0}^{\infty }du\,\frac{u^{D-1}}{%
c_{1}(u)c_{2}(u)e^{2u}-1}  \notag \\
&&\times \left\{ 2Z_{\nu }(\eta ,u)+\left[ Z_{\nu }(\eta ,u)+b_{\nu }\left(
u/z_{0}\right) ^{2}Z(\eta ,u)\right]
\sum_{j=1,2}c_{j}(u)e^{2u|z-z_{j}|/z_{0}}\right\} \,,  \label{Tnu2}
\end{eqnarray}%
with the functions%
\begin{eqnarray}
Z_{0}(\eta ,u) &=&\widehat{P}_{0}Z(\eta ,u)-u^{2}Y(\eta ,u)/z_{0}^{2},
\notag \\
Z_{l}(\eta ,u) &=&\frac{u^{2}/z_{0}^{2}}{D-1}Y(\eta ,u)-\left( \widehat{P}%
_{1}+\frac{u^{2}/z_{0}^{2}}{D-1}\right) Z(\eta ,u),  \notag \\
Z_{D}(\eta ,u) &=&\left( u^{2}/z_{0}^{2}-\widehat{P}_{1}\right) Z(\eta ,u),
\label{ZDn}
\end{eqnarray}%
and%
\begin{equation}
Y(\eta ,u)=\int_{0}^{1}ds\,s^{3}(1-s^{2})^{(D-3)/2}U(\eta ,us).  \label{Y}
\end{equation}%
For the off-diagonal component we find%
\begin{eqnarray}
\langle T_{0}^{D}\rangle &=&-\frac{A_{D}}{z_{0}^{D}a^{2}}\int_{0}^{\infty
}du\,u^{D}\,\frac{\sum_{j=1,2}c_{j}(u)(-1)^{j-1}e^{2u|z-z_{j}|/z_{0}}}{%
c_{1}(u)c_{2}(u)e^{2u}-1}  \notag \\
&&\times \left[ \left( 1/2-2\xi \right) \partial _{\eta }+2\xi \tilde{H}%
\right] Z(\eta ,u).  \label{T0Dn}
\end{eqnarray}%
The dependence of the VEVs on the background geometry enters through the
functions $Z(\eta ,u)$ and $Y(\eta ,u)$.

In the regions $z<z_{1}$ and $z>z_{2}$, the alternative expressions for the
VEVs are given by
\begin{eqnarray}
\langle T_{\nu }^{\nu }\rangle _{j} &=&\langle T_{\nu }^{\nu }\rangle _{0}+%
\frac{A_{D}}{a^{2}}\int_{0}^{\infty }du\,u^{D-1}\frac{\beta _{j}u+1}{\beta
_{j}u-1}\frac{Z_{\nu }(\eta ,uz_{0})+b_{\nu }u^{2}Z(\eta ,uz_{0})}{%
e^{2u|z-z_{j}|}},  \notag \\
\langle T_{0}^{D}\rangle _{j} &=&-\frac{\left( -1\right) ^{j}A_{D}}{a^{2}}%
\int_{0}^{\infty }du\,u^{D}\frac{\beta _{j}u+1}{\beta _{j}u-1}\frac{\left[
\left( 1/2-2\xi \right) \partial _{\eta }+2\xi \tilde{H}\right] Z(\eta
,uz_{0})}{e^{2u|z-z_{j}|}},  \label{T0D1pl2}
\end{eqnarray}%
with $j=1$ and $j=2$, respectively. Note that for the Minkowski bulk the
normal stress in the geometry of a single plate vanishes.

Under the adiabatic condition (\ref{fadiab}), we can find simple asymptotic
expressions of the VEVs near the boundaries for general case of the scale
factor. By taking into account that the dominant contribution to the
integral in (\ref{Tnu2}) comes from large values of $u$ and using the
asymptotic expression (\ref{Zas}), near the plate at $z=z_{j}$, to the
leading order one finds (no summation over $\nu $)%
\begin{equation}
\langle T_{\nu }^{\nu }\rangle \approx \left( 2\delta _{0\beta
_{j}}-1\right) \frac{D\Gamma ((D+1)/2)(\xi -\xi _{D})}{2^{D}\pi
^{(D+1)/2}\left( a|z-z_{j}|\right) ^{D+1}},  \label{TnuNear}
\end{equation}%
for $\nu =0,1,\ldots ,D-1$. For the normal stress the leading term vanishes
and it is needed to keep the next-to-leading term. It is more convenient to
find the corresponding asymptotic expression by using the second equation in
(\ref{ConsEq}) and the asymptotic expression for the energy flux. For the
latter from (\ref{T0Dn}) and (\ref{Zas}) we get%
\begin{equation}
\langle T_{0}^{D}\rangle \approx \left( 2\delta _{0\beta _{j}}-1\right)
\frac{2\left( -1\right) ^{j}D\left( \xi -\xi _{D}\right) H}{(4\pi
)^{(D+1)/2}\left( a|z-z_{j}|\right) ^{D}}\Gamma ((D+1)/2).  \label{T0DNear}
\end{equation}%
Combining this with (\ref{ConsEq}), one obtains the asymptotic for the
normal stress:%
\begin{equation}
\langle T_{D}^{D}\rangle \approx \left( 1-2\delta _{0\beta _{j}}\right)
\frac{D\left( \xi -\xi _{D}\right) \Gamma ((D-1)/2)}{(4\pi )^{(D+1)/2}\left(
a|z-z_{j}|\right) ^{D-1}}\frac{\ddot{a}}{a}.  \label{TDDNear}
\end{equation}%
The leading terms in the near-plate asymptotic expansions for the diagonal
components with $\nu \neq D$, given by (\ref{TnuNear}), coincide with the
corresponding expressions in the Minkowski bulk, with the distance $|z-z_{j}|
$ replaced by the proper distance $a(\eta )|z-z_{j}|$. For the Minkowski
bulk, the normal stress $\langle T_{D}^{D}\rangle $ does not depend on the
coordinate $z$. This property is already seen from the second equation in (%
\ref{ConsEq}), by taking into account that in the Minkowski bulk $\langle
T_{0}^{D}\rangle =0$. Hence, we see that the cosmological expansion
essentially changes the behavior of the normal stress. In particular, near
the plates the normal stress has different signs for accelerated and
deccelerated expansions. Eqs. (\ref{TnuNear})-(\ref{TDDNear}) present the
leading-order terms in the asymptotic expansions of the VEVs over the
distance from the plate $z=z_{j}$. These leading terms do not depend on the
field mass and vanish for a conformally coupled field. The next-to-leading
order terms, in general, will depend on the mass they do not vanish in the
conformally coupled case.

As is seen from Eqs. (\ref{TnuNear})-(\ref{TDDNear}), the VEV of the
energy-momentum tensor diverges on the boundaries. These types of
divergences are well known in quantum field theory with boundaries and they
have been investigated for various bulk and boundary geometries. For
cosmological backgrounds, an essential difference from the corresponding
problem in the Minkowski bulk is that the normal stress diverges on the
boundary. For the Minkowski bulk it remains finite everywhere. Moreover, the
corresponding VEV does not depend on $z$ in the region between the plates
and vanishes in the regions $z<z_{1}$ and $z>z_{2}$. From Eqs. (\ref{TnuNear}%
)-(\ref{TDDNear}) it follows that near the plates the VEVs for a field with $%
\xi \neq \xi _{D}$ have opposite signs for Dirichlet ($\beta _{j}=0$) and
non-Dirichlet boundary conditions.

On the base of the results given above, we can investigate the vacuum
densities induced by a thick domain wall in the background of FRW spacetime.
This is done in the way similar to that used in \cite{Saha07} for a thick
brane on the anti-de Sitter bulk. For a thick domain wall with the thickness
$2b$, we write the line element for the interior geometry in the form $%
ds^{2}=a^{2}(\eta )[e^{u(z)}d\eta ^{2}-e^{v(z)}d\mathbf{x}_{\parallel
}^{2}-e^{w(z)}dz^{2}]$, $|z|<b$. In the regions $|z|>b$, the line element is
given by (\ref{ds2}). The functions $u(z)$, $v(z)$ and $w(z)$ are continuous
on the boundaries $z=-b$ and $z=b$. For the symmetric domain wall these
functions are even functions of $z$. It can be shown that (the details will
be presented elsewhere) the VEVs in the region $z>b$ are given by the
expressions (\ref{phi22}) and (\ref{T0D1pl}) with $z_{j}=b$ and with the
Robin coefficient $\beta _{j}$ being a function of the quantum numbers $k$
and $\lambda $. This function is determined by the matching conditions for
the scalar field modes in the interior and exterior regions.

\subsection{The Casimir force}

In the geometry of a single plate the vacuum pressures on the right- and
left-hand sides of the plate compensate each other and the corresponding net
force is zero. Consequently, for the two plates geometry, the resulting
force per unit surface is determined by the part in the normal stress $%
\langle T_{D}^{D}\rangle $ induced by the second plate:%
\begin{equation}
P_{j}=-\left( \langle T_{D}^{D}\rangle -\langle T_{D}^{D}\rangle _{j}\right)
|_{z=z_{j}},  \label{Pj0}
\end{equation}%
where $\langle T_{D}^{D}\rangle $ is the normal stress in the region between
the plates. By taking into account the expressions given above, we get%
\begin{equation}
P_{j}=\frac{A_{D}}{z_{0}a^{2}}\int_{0}^{\infty }dk\,k^{D-2}\int_{0}^{\infty
}du\frac{\left[ 2+c_{j}(u)+1/c_{j}(u)\right] \hat{P}_{1}-2u^{2}/z_{0}^{2}}{%
c_{1}(u)c_{2}(u)e^{2u}-1}W(\eta ,\eta ,u,k).  \label{Pj}
\end{equation}%
The force is attractive for $P_{j}<0$ and repulsive for $P_{j}>0$. In the
problem on the Minkowski bulk one has $\hat{P}_{1}W(\eta ,\eta ,u,k)=0$ and
the first term in the numerator of the integrand in (\ref{Pj}) vanishes.
Hence, the Casimir force for the Minkowski bulk is the same for both the
plates, regardless of the values of the coefficients in the Robin boundary
conditions. This is not the case for general FRW spacetime.

An alternative representation for the Casimir force is obtained by using the
expression (\ref{Tnu2}) for the normal stress:%
\begin{equation}
P_{j}=\frac{A_{D}}{z_{0}^{D}a^{2}}\int_{0}^{\infty }du\,u^{D-1}\frac{\left[
2+c_{j}(u)+1/c_{j}(u)\right] \hat{P}_{1}-2u^{2}/z_{0}^{2}}{%
c_{1}(u)c_{2}(u)e^{2u}-1}Z(\eta ,u),  \label{Pj2}
\end{equation}%
with the function $Z(\eta ,u)$ defined by (\ref{Z}). Depending on the Robin
coefficients and on the vacuum state, the forces corresponding to (\ref{Pj2}%
) can be either attractive or repulsive. In particular, one can have the
situation when the forces are repulsive at small separations between the
plates and attractive at large separation.

Assuming that the scalar modes satisfy the adiabatic condition with the
small wavelength asymptotic (\ref{fadiab}), we can find the asymptotic of
the Casimir force at small separation between the plates. Under the
assumption $1/(az_{0})\gg m,\sqrt{|\dot{H}|},H$, the dominant contribution
in (\ref{Pj2}) comes from the second term in the numerator of the integrand.
By using the asymptotic (\ref{Zas}), to the leading order one gets%
\begin{equation}
P_{j}\approx -\frac{2(4\pi )^{-D/2}}{(z_{0}a)^{D+1}\Gamma (D/2)}%
\int_{0}^{\infty }du\,\frac{u^{D}}{c_{1}(u)c_{2}(u)e^{2u}-1}.  \label{PjNear}
\end{equation}%
The expression in the right-hand side coincides with the Casimir pressure
for the plates in the Minkowski spacetime for a massless scalar field.

\section{Special cases}

\label{sec:Spec}

In this section we consider some special cases of the general results given
above. For the Minkowski bulk $a(t)=1$ and for the modes realizing the
standard Minkwoski vacuum one has $f(\eta ,\gamma )=e^{-i\omega \eta }/\sqrt{%
2\omega }$ with $\omega =\sqrt{\gamma ^{2}+m^{2}}$, and $w(\eta ,\eta
,\gamma )=1/\omega $. From here it follows that $U(\eta ,z_{0}x)=0$ for $x<m$
and $U(\eta ,z_{0}x)=2/\sqrt{x^{2}-m^{2}}$ for $x>m$. For the function
appearing in the expressions (\ref{phi2n}), (\ref{Tnu}) and (\ref{Pj}) one
gets $W(\eta ,\eta ,u,k)=0$ for $u<z_{0}\sqrt{k^{2}+m^{2}}$ and%
\begin{equation}
W(\eta ,\eta ,u,k)=\frac{2}{\sqrt{u^{2}/z_{0}^{2}-k^{2}-m^{2}}}
\label{WMink}
\end{equation}
for $u>z_{0}\sqrt{k^{2}+m^{2}}$. In this special case, the function $Z(\eta
,u)$ in the expressions for the VEVs is simplified to%
\begin{equation}
Z(\eta ,u)=\frac{\sqrt{\pi }\Gamma ((D-1)/2)z_{0}}{\Gamma (D/2)u}\left[
1-\left( z_{0}m/u\right) ^{2}\right] ^{D/2-1},  \label{ZMink}
\end{equation}%
for $u\geqslant z_{0}m$ and $Z(\eta ,u)=0$ for $u<z_{0}m$. Substituting the
expression (\ref{ZMink}) into the general formulas given above, we obtain
the VEVs for the Robin plates in Minkowski spacetime (see \cite{RS02} for
the VEVs in the massless case and \cite{Saha06} for a massive scalar field.
Note that in \cite{Saha06} the VEVs in the Minkowski bulk are obtained as a
limiting case of the corresponding problem with two uniformly accelerated
plates moving through the Fulling-Rindler vacuum state).

\subsection{Conformally coupled massless field}

For a conformally coupled massless field one has $\xi =\xi _{D}$ and $m=0$.
As it follows from (\ref{g1}), the general solution for the function $f(\eta
,\gamma )$ has the form
\begin{equation}
f(\eta ,\gamma )=\frac{a^{(1-D)/2}}{\sqrt{2\gamma }}\left( c_{1}e^{-i\gamma
\eta }+c_{2}e^{i\gamma \eta }\right) ,  \label{fcc}
\end{equation}%
where the factor $1/\sqrt{2\gamma }$ is extracted for the further
convenience. One of the coefficients is determined by the normalization
condition whereas the second one is fixed by the choice of the vacuum state.
As a vacuum state we will take the state corresponding to the standard
Minkowskian vacuum in the adiabatic limit $a(\eta )=\mathrm{const}$. This
corresponds to the choice $c_{2}=0$ and from the normalization condition (%
\ref{cond2}) one gets $|c_{1}|^{2}=1$.

For the function $W(\eta ,\eta ^{\prime },u,k)$ in the expressions of the
VEVs we find $W(\eta ,\eta ^{\prime },u,k)=0$ in the region $u<z_{0}k$ and
\begin{equation}
W(\eta ,\eta ^{\prime },u,k)=2a^{1-D}\frac{\cosh [(\eta -\eta ^{\prime })%
\sqrt{u^{2}/z_{0}^{2}-k^{2}}]}{\sqrt{u^{2}/z_{0}^{2}-k^{2}}},  \label{Wcc}
\end{equation}%
for $u>z_{0}k$ and, hence, $U(\eta ,x)=2a^{1-D}z_{0}/x$. From (\ref{phi23}),
for the VEV of the field squared one finds
\begin{equation}
\langle \phi ^{2}\rangle =\langle \phi ^{2}\rangle _{0}+\frac{\left(
az_{0}\right) ^{1-D}}{(4\pi )^{D/2}\Gamma (D/2)}\,\int_{0}^{\infty
}du\,u^{D-2}\frac{2+\sum_{j=1,2}c_{j}(u)e^{2u|z-z_{j}|/z_{0}}}{%
c_{1}(u)c_{2}(u)e^{2u}-1}.  \label{phi2cc}
\end{equation}%
In a similar way, we can see that the VEV of the energy-momentum tensor
takes the form%
\begin{equation}
\langle T_{\nu }^{\mu }\rangle =\langle T_{\nu }^{\mu }\rangle _{0}-\frac{%
\mathrm{diag}(1,1,\ldots ,1,-D)}{(4\pi )^{D/2}\Gamma (D/2+1)(az_{0})^{D+1}}%
\int_{0}^{\infty }du\,\frac{u^{D}}{c_{1}(u)c_{2}(u)e^{2u}-1}.  \label{Tmucc}
\end{equation}%
In this special case the vacuum energy-momentum tensor is spatially
homogeneous and diagonal. Of course, the boundary-induced contributions in (%
\ref{phi2cc}) and (\ref{Tmucc}) could be directly obtained from the
corresponding expressions in the Minkowski bulk by using the standard result
for conformally related problems (see, for instance, \cite{Birr82B}).

\subsection{de Sitter bulk}

As a next application we consider the case of de Sitter bulk with $%
a(t)=e^{Ht}$, $H=\mathrm{const}$ (the renormalized expectation value of the
energy-momentum tensor for an arbitrary homogeneous and isotropic physical
initial state of a scalar field in de Sitter spacetime, in the absence of
boundaries, has been investigated in \cite{Ande00}). The corresponding scale
factor in conformal time has the form $a(\eta )=-1/(H\eta )$ with $-\infty
<\eta \leqslant 0$. In this case one has $\tilde{H}=-1/\eta $ and $%
R=D(D+1)H^{2}$. The general solution of the equation (\ref{f1}) is the
linear combination of the functions $|\eta |^{D/2}H_{\nu }^{(1)}(\gamma
|\eta |)$ and $|\eta |^{D/2}H_{\nu }^{(2)}(\gamma |\eta |)$, with $H_{\nu
}^{(1,2)}(x)$ being the Hankel functions and
\begin{equation}
\nu =\sqrt{D^{2}-4D(D+1)\xi -m^{2}/H^{2}}.  \label{nu}
\end{equation}%
For the further convenience we write the Hankel functions in terms of the
Macdonald function $K_{\nu }(x)$ \cite{Abra72}:%
\begin{equation}
f(\eta ,\gamma )=\frac{|\eta |^{D/2}}{\sqrt{\pi }\alpha ^{(D-1)/2}}\left[
d_{1}K_{\nu }(\gamma |\eta |e^{-\pi i/2})+d_{2}K_{\nu }(\gamma |\eta |e^{\pi
i/2})\right] ,  \label{fds}
\end{equation}%
where the parameter $\nu $ is either positive or purely imaginary. From the
condition (\ref{cond2}) we get the relation%
\begin{equation}
|d_{1}|^{2}-|d_{2}|^{2}=1,  \label{drel}
\end{equation}%
between the coefficients.

By using the relation \cite{Abra72}%
\begin{equation}
K_{\nu }(\gamma |\eta |e^{\pm \pi i})=e^{\mp \nu \pi i}K_{\nu }(\gamma |\eta
|)\mp \pi iI_{\nu }(\gamma |\eta |),  \label{Krel}
\end{equation}%
for the function appearing in the expressions of the VEVs we get $W(\eta
,\eta ^{\prime },u,k)=0$ for $u<kz_{0}$ and%
\begin{eqnarray}
&&W(\eta ,\eta ^{\prime },u,k)=2H^{D-1}|\eta \eta ^{\prime }|^{D/2}\left\{
i\pi \left( d_{1}d_{2}^{\ast }-d_{1}^{\ast }d_{2}\right) I_{\nu }(y)I_{\nu
}(y^{\prime })\right.  \notag \\
&&\qquad \left. +\left( \left\vert d_{1}\right\vert ^{2}+\left\vert
d_{2}\right\vert ^{2}+e^{\nu \pi i}d_{1}d_{2}^{\ast }+e^{-\nu \pi
i}d_{1}^{\ast }d_{2}\right) \left[ I_{-\nu }(y)K_{\nu }(y^{\prime })+I_{\nu
}(y^{\prime })K_{\nu }(y)\right] \right\} ,  \label{Wds}
\end{eqnarray}%
for $u>kz_{0}$, where%
\begin{equation}
y=|\eta |\sqrt{u^{2}/z_{0}^{2}-k^{2}},\;y^{\prime }=|\eta ^{\prime }|\sqrt{%
u^{2}/z_{0}^{2}-k^{2}}.  \label{yy}
\end{equation}%
For the function in (\ref{Wds}) one has
\begin{equation}
I_{-\nu }(y)K_{\nu }(y^{\prime })+I_{\nu }(y^{\prime })K_{\nu }(y)=-\frac{%
\pi }{2}\frac{I_{\nu }(y)I_{\nu }(y^{\prime })-I_{-\nu }(y)I_{-\nu
}(y^{\prime })}{\sin \left( \nu \pi \right) },  \label{RelFunc}
\end{equation}%
which shows that this function is real for both the real and purely
imaginary values for $\nu $.

Note that in the expressions of the VEVs only the relative phase of the
coefficients $d_{1}$ and $d_{2}$ is relevant and, hence, by taking into
account the relation (\ref{drel}), we can take the parametrization%
\begin{equation}
d_{1}=\cosh \alpha ,\;d_{2}=e^{i\beta }\sinh \alpha ,  \label{c12par}
\end{equation}%
in terms of new real parameters $\alpha $ and $0\leqslant \beta <2\pi $.
With this parametrization, for the function (\ref{Wds}) one gets%
\begin{eqnarray}
W(\eta ,\eta ^{\prime },u,k) &=&2H^{D-1}|\eta \eta ^{\prime }|^{D/2}\left\{
\pi \sinh (2\alpha )\sin \beta \,I_{\nu }(y)I_{\nu }(y^{\prime })\right.
\notag \\
&&+\left[ \cosh (2\alpha )+\sinh (2\alpha )\cos \left( \beta -\nu \pi
\right) \right]  \notag \\
&&\left. \times \left[ I_{-\nu }(y)K_{\nu }(y^{\prime })+I_{\nu }(y^{\prime
})K_{\nu }(y)\right] \right\} .  \label{Wds2}
\end{eqnarray}%
The modes (\ref{fds}) correspond to the two-parameter $\left( \alpha ,\beta
\right) $ family of vacuum states in de Sitter spacetime. As it has been
discussed in \cite{Alle85}, in the absence of the plates the de Sitter
invariant vacuum states correspond to $\beta =0$. The Bunch-Davies (or
Euclidean) vacuum state \cite{Bunc78} is a special case of de Sitter
invariant vacua and corresponds to $\alpha =0$. In general, one has a
one-parameter family of de Sitter invariant vacuum states specified by the
parameter $\alpha $ ($\alpha $-states or $\alpha $-vacua in de Sitter space,
for the discussion of the role of these states in inflationary models see,
for example, \cite{Kalo02}).

The transformations of the boundary-induced contributions in the VEVs, we
have described above, are valid for dS invariant vacua only. In this special
case the function (\ref{Wds2}) takes the form%
\begin{equation}
W(\eta ,\eta ^{\prime },u,k)=2H^{D-1}b(\alpha )|\eta \eta ^{\prime }|^{D/2}
\left[ I_{-\nu }(y)K_{\nu }(y^{\prime })+I_{\nu }(y^{\prime })K_{\nu }(y)%
\right] ,  \label{WBD}
\end{equation}%
with%
\begin{equation}
b(\alpha )=\cosh (2\alpha )+\sinh (2\alpha )\cos \left( \nu \pi \right) .
\label{balf}
\end{equation}%
From here it follows that the VEVs of the field squared and of the
energy-momentum tensor for a dS invariant vacuum state with a given $\alpha $
are obtained from the corresponding VEVs\ in the Bunch-Davies vacuum state,
investigated in \cite{Saha09}, multiplying by the factor $b(\alpha )$. For
real values of the parameter $\nu $ this factor is always positive. For
purely imaginary $\nu $, the factor $b(\alpha )$ can be negative. In this
case, compared with the Bunch-Davies vacuum state, the Casimir forces for
the corresponding $\alpha $-vacuum change the sign.

\section{Conclusion}

\label{sec:Conc}

We have studied the scalar Casimir effect for the geometry of two parallel
plates on the spatially flat FRW background for a general case of the scale
factor. On the plates the field obeys the Robin boundary conditions (\ref{6}%
) with the coefficients proportional to the scale factor. In the model under
consideration, all the properties of the vacuum state are encoded in
two-point functions and, as the first step in the investigation of the VEVs
for physical observables bilinear in the field operator, we have evaluated
the Hadamard function. By using the Abel-Plana-type summation formula for
the eigenvalues of the quantum number $\lambda $, the boundary-induced
contribution is explicitly extracted. This contribution in the geometry of a
single plate and in the region between two plates is given by the last terms
in (\ref{Gj1}) and (\ref{Had3}), respectively. In the corresponding
evaluation we have not fixed the vacuum state. In order to specify the
vacuum state, an additional condition should be imposed on the function $%
f(\eta ,\gamma )$ appearing in the expression (\ref{phi2}) for the scalar
modes. In particular, for the modes obeying the adiabatic condition this
function has the small wavelength asymptotic (\ref{fadiab}). In the limit of
a constant scale factor, these modes approach the positive energy solutions
used for the quantization of a scalar field in the Minkowski bulk.

As important local characteristics of the vacuum state, we have considered
the VEVs of the field squared and of the energy-momentum tensor. The VEV of
the field squared is given by the expression (\ref{phi23}) in the region
between the plates and by (\ref{phi24}) in the regions $z<z_{1}$ and $%
z>z_{2} $. For points away from the boundaries the renormalization is
reduced to that for the boundary free-part $\langle \phi ^{2}\rangle _{0}$.
The information on the background geometry is encoded in the function $%
Z(\eta ,u) $, defined by the relation (\ref{Z}). For the vacuum state
realized by the modes obeying the adiabatic condition, the leading term in
asymptotic expansion of the field squared near the plates is given by the
expression (\ref{phi2near}). It is obtained from the corresponding
asymptotic in the problem on Minkowski bulk replacing the distance from the
plate by the proper distance. Near the plates the dominant contribution to
the VEVs come from the modes with small wavelengths, the influence of the
gravitational field on which is weak.

The diagonal components of the VEV of the energy-momentum tensor in the
region between the plates are given by the formula (\ref{Tnu2}), where the
functions in the boundary-induced contribution are defined by (\ref{ZDn}).
Unlike to the case of the Minkowski bulk, the corresponding normal stress is
inhomogeneous. Another feature of the Casimir effect in the expanding bulk
is the presence of the nonzero energy flux along the direction normal to the
plates. This flux is described by the off-diagonal component of the vacuum
energy-momentum tensor, given by the expression (\ref{T0Dn}). Depending on
the Robin coefficients and on the vacuum state, the flux can be either
positive or negative. For boundaries with the same Robin coefficients, the
energy flux vanishes on the plane $z=(z_{1}+z_{2})/2$ and has opposite signs
in the right-hand and left-hand regions with respect to this plane. In the
regions $z<z_{1}$ and $z>z_{2}$ the vacuum energy-momentum tensor coincides
with that for the geometry of a single plate and is given by the formulas (%
\ref{T0D1pl2}). The corresponding normal stress and the energy flux vanish
in the Minkowskian limit. Under the adiabatic condition for the scalar
modes, the leading term in the near-plate expansion of the diagonal
components $\langle T_{\nu }^{\nu }\rangle $ for $\nu \neq D$ coincides with
that for plates in the Minkowski spacetime. For the energy flux and the
normal stress the corresponding asymptotics are given by the expressions (%
\ref{T0DNear}) and (\ref{TDDNear}). In particular, for the normal stress the
asymptotic behavior on the FRW bulk is completely different from that for
the Minkowski spacetime. In the latter case the normal stress is finite on
the plates.

The Casimir force per unit surface of the plate at $z=z_{j}$ is determined
by the expression (\ref{Pj}). An important difference from the corresponding
result in the Minkowski bulk is that for a scalar field with $\beta _{1}\neq
\beta _{2}$ the forces acting on the right and left plates, in general, are
different. Depending on the Robin coefficients and on the vacuum state under
consideration, these forces can be either attractive or repulsive. Assuming
that the modes used in the quantization procedure obey the adiabatic
condition, for the leading term in the asymptotic expansion of the Casimir
force at small distances between the plates one gets the expression (\ref%
{PjNear}).

In Section \ref{sec:Spec}, two special cases of general results are
discussed. In the first example we have considered a conformally coupled
massless field assuming that the field is prepared in the vacuum state that
corresponds to the Minkowskian vacuum in the adiabatic limit. In this case,
the boundary-induced contributions to the VEVs of the field squared and of
the energy-momentum tensor are obtained from the corresponding VEVs in the
Minkowski bulk by using the standard relation for conformally coupled
problems. In particular, the vacuum energy-momentum tensor is diagonal. In
the second example, the de Sitter spacetime is considered as a background
geometry. For this geometry one has a one-parameter family of the de Sitter
invariant vacuum states specified by the real parameter $\alpha $. The
corresponding function $W(\eta ,\eta ^{\prime },u,k)$, appearing in the
expressions for the VEVs, is given by the expression (\ref{WBD}) with the
coefficient $b(\alpha )$ defined by (\ref{balf}). In the special case $%
\alpha =0$, we obtain the results for the Bunch-Davies vacuum state
previously discussed in the literature. For imaginary values of the
parameter $\nu $, depending on the parameter $\alpha $, the Casimir forces
for the $\alpha $-vacua may have opposite signs compared with the
Bunch-Davies vacuum.

\section*{Acknowledgments}

E.R.B.M. thanks Conselho Nacional de Desenvolvimento Cient\'{\i}fico e Tecnol%
\'{o}gico (CNPq) for the partial financial support by the Project No.
313137/2014-5. A.A.S. was supported by the State Committee of Science
Ministry of Education and Science RA, within the frame of Grant No. SCS
15T-1C110, and by the Armenian National Science and Education Fund (ANSEF)
Grant No. hepth-4172. The work of M.R.S. has been supported by the Research
Institute for Astronomy and Astrophysics of Maragha (RIAAM).

\end{document}